\documentclass[aps,prl,twocolumn,groupedaddress]{revtex4}
\usepackage{graphicx}
\usepackage{amssymb}
\usepackage{bm}
\usepackage{natbib}

\newcommand{\etal}{\textit{et al.}}

\begin{document}


\title{Bosons in Disordered Optical Potentials}
\author{Pearl J.Y. Louis and Makoto Tsubota}
\affiliation{Department of Physics, Osaka City University, Sugimoto 3-3-138, Sumiyoshi-ku, Osaka 558-8585, Japan}


\begin{abstract}
In this work we systematically investigate the condensate properties, superfluid properties and quantum phase transitions in interacting Bose gases trapped in disordered optical potentials.  We numerically solve the Bose-Hubbard Hamiltonian exactly for different: (a) types of disorder, (b) disorder strengths, and (c) interatomic interactions.  The three types of disorder studied are: quasiperiodic disorder, uniform random disorder and random speckle-type disorder.  We find that the Bose glass, as identified by Fisher~\etal\ [Phys. Rev. B {\bf 40}, 546 (1989)], contains a normal condensate component and we show how the three different factors listed above affect it.
\end{abstract}

\maketitle



The properties of Bose systems can be strongly affected by interatomic interactions and disorder.  In the noninteracting case disorder leads to the insulating Anderson localization~\cite{Anderson}. If interatomic interactions are also present, the interplay between the interactions and disorder can lead to a rich and complex arrangement of different quantum phases~\cite{FisherWeichman}.  It is even possible for states to develop where the BEC (Bose-Einstein condensate) fraction is non-superfluid~\cite{HuangMeng,GiorginiPitaevskii}.  

Our work has two major goals.  The first is to show that the normal fluid BEC found by Huang and Meng~\cite{HuangMeng} and Giorgini~\etal~\cite{GiorginiPitaevskii} can be identified with the Bose glass insulating state found by Fisher~\etal~\cite{FisherWeichman}.  Using the tight-binding Bose-Hubbard lattice model, Fisher~\etal~\cite{FisherWeichman} were able to describe an insulating state in disordered lattices which exists due to the cooperative effect of repulsive interactions and disorder.  This `Bose glass' (BG) is gapless and compressible. Taking an alternative approach with a microscopic low-density model involving a dilute low-temperature hard-sphere gas with random scatterers in the limit of weak disorder, Huang and Meng~\cite{HuangMeng} and Giorgini~\etal~\cite{GiorginiPitaevskii} found that the superfluid is more depleted by the disorder than the condensate. That is, there exists \emph{BEC without superfluidity}.  

It was shown previously using the disordered Bose-Hubbard model that it is possible to obtain a normal condensate fraction for certain parameters~\cite{DamskiZakrzewski} and that there is a normal condensate fraction in the Bose glass phase for quasiperiodic disorder~\cite{RothBurnett}.  In this work we show that the parameter regime where this occurs can be identified with the BG region described by Fisher~\etal~\cite{FisherWeichman} by showing that the BG phase and the normal condensate phase occur together as we change the interaction strength, disorder strength and the \emph{type} of disorder.  We also find that in the BG regime the spatial correlations go to zero with distance but remain significant over multiple lattice sites.

Our second goal is to show how the properties of `dirty' Bose gases change with: (a) the type of disorder, (b) the disorder strength, and (c) the interatomic interaction strength, especially with respect to the existence of condensate without superfluidity.  We show that there are significant differences in the phase diagram as we change the type of disorder and the nature of the differences depend on the type of disorder.

To date the majority of the work on `dirty' bosons has focused on liquid $^{4}$He in porous materials such as Vycor glass or aerogel~\cite{Reppy}. In recent years however, weakly-interacting dilute Bose gases, especially those trapped in optical dipole potentials, have become increasingly important in this field~\cite{LyeFallani,ClementVaron,SchulteDrenkelforth,FallaniLye}.  An important advantage of these systems is their flexibility and their ease of control over important experimental parameters. This can be seen in recent experiments in which a variety of different types of disordered optical potentials of controllable strength were studied e.g., quasiperiodic lattices~\cite{FallaniLye}, speckle potentials~\cite{LyeFallani,ClementVaron}, and lattices with superimposed disorder~\cite{SchulteDrenkelforth}.  These recent experimental developments make an exploration of how condensate and superfluid properties change with the properties of the disorder increasingly relevant.  Another advantage of these systems over $^4$He is that BEC without superfluidity only occurs in very dilute systems~\cite{AstrakharchikBoronat}.


The many-body quantum state for {\it N} interacting bosons in a 1D lattice with $N_{s}$ lattice sites can be described using the 1D Bose-Hubbard (BH) Hamiltonian,
\begin{equation}\label{eqBHH}
\hat{H}=-J\sum_{i=1}^{N_s}(\hat{a}_{i+1}^{\dagger}\hat{a}_{i}+\mathrm{H.c.})+\frac{U}{2}\sum_{i=1}^{N_s}\hat{n}_{i}(\hat{n}_{i}-1)+\sum_{i=1}^{N_s}\varepsilon_{i}\hat{n}_{i}.
\end{equation} 
$\hat{a}_{i}^{\dagger}$, $\hat{a}_{i}$ are the boson creation and annihilation operators respectively at the {\it i-}th lattice site.  $\hat{n}_{i}=\hat{a}_{i}^{\dagger}\hat{a}_{i}$.  {\it J} is the strength of tunnelling between neighboring sites and {\it U} the strength of the on-site interatomic interactions.  The values of {\it J} and {\it U} can be calculated as in Ref.~\cite{JakschBruder}.  $\varepsilon_{i}$ is the on-site single particle energy at site {\it i}.

The BH Hamiltonian~(\ref{eqBHH}) can be solved numerically by diagonalization in the occupation number basis.  The large size of this basis, \mbox{$D=(N+N_s-1)!/(N!(N_s-1)!)$} restricts the $N_s$ and $N$ that can be handled using diagonalization. However, diagonalization solves Eq.~(\ref{eqBHH}) in a numerically exact and transparent way.  As will be seen, even in our finite systems, significant differences can be seen as the type of disorder, the disorder strength and the interaction strength change. 

Significantly, we do not make the assumption that the condensate fraction, $f_c$ and the superfluid fraction, $f_s$ are the same but instead define the two separately following the definitions used in~\cite{RothBurnett} and~\cite{DamskiZakrzewski}.  $f_{s}$ can be calculated from the response of the system to a very small phase variation, $\Delta\theta(x)=\Theta$, across the lattice.  As long as $\Theta$ is small, the increase in energy is due solely to the kinetic energy of the superflow.  Assuming that the phase variation is linear,
\begin{equation}\label{eqSF}
f_{s}=\frac{N_s^{2}}{NJ}\frac{E_{0}(\Delta\theta=\Theta)-E_{0}(\Delta\theta=0)}{\Theta^{2}}
\end{equation}
in terms of the BH coefficients~\cite{RothBurnett,DamskiZakrzewski}.  Instead of using twisted boundary conditions, it is easier to use a unitary transformation to make a `twisted Hamiltonian' and use periodic boundary conditions.

However BEC is not defined in terms of transport properties but as pointed out by Penrose and Onsager from the properties of the one-body density matrix~\cite{PenroseOnsager}.  Following this, $f_{c}$ is taken to be the largest eigenvalue of the single-particle density matrix for the ground state $|\psi_{0}\rangle$, $\rho_{ij}=\langle\psi_{0}|\hat{a}_{j}^{\dagger}\hat{a}_{i}|\psi_{0}\rangle$, divided by $N$, the total number of atoms.  However, because of finite size effects, the restriction $f_{c}>1/N_s$ exists. Due to the 1D nature of the system this is not a true condensate.  However the physical similarity between the definition of $f_{c}$ here and true condensate and the finite size of the system means that it should still be useful to consider what happens in 1D.

\begin{figure}
\includegraphics[width=8.6cm]{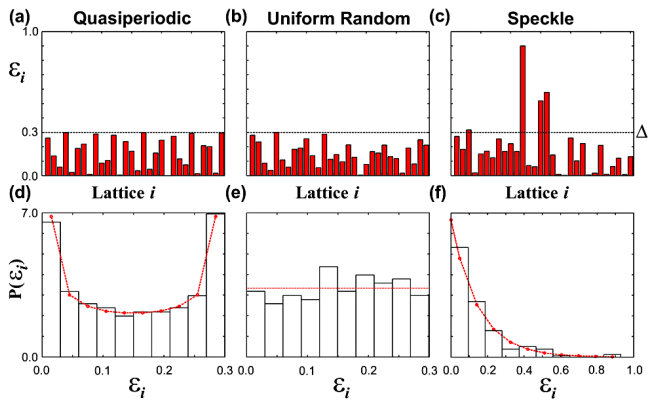}
\caption{\label{figPotSamples} (Color online) The on-site single particle energies $\varepsilon_{i}$ for: (a) The quasiperiodic lattice $\varepsilon_{i}=\Delta\sin^{2}(kx_{i})$ for $\Delta=0.3$, \mbox{$k=(1+\sqrt{5})/2$} and $x_i=i\pi$.  (b) Uniform random disorder for $\Delta=0.3$.  (c) Speckle disorder with standard deviation $\sigma_{SP}$; \mbox{$\Delta=2\sigma_{SP}=0.3$}.  30 lattice sites are shown. (d), (e) and (f): The probability distributions $P(\varepsilon_{i})$ for lattices (a), (b) and (c) respectively.  The histograms show $P(\varepsilon_{i})$ over 170 lattice sites and the dotted line shows the expected result over many realizations.}
\end{figure}

We compare three different types of disordered lattices: quasiperiodic~\cite{FallaniLye}, speckle ~\cite{LyeFallani,ClementVaron} and uniform random ~\cite{FisherWeichman}. It is assumed that the lattice is deep enough that disorder with strength $\Delta$ \emph{only} alters $\varepsilon_{i}$ significantly~\cite{DamskiZakrzewski}.  $\Delta=0$ obviously represents a homogeneous single-periodic lattice.  For uniform random disorder $\varepsilon_{i}$ varies flatly as $0\leq\varepsilon_{i}\leq\Delta$ [see Figs.~\ref{figPotSamples}(b) and (e)].

The definition of $\Delta$ for speckle-type disorder is different as there is no upper bound for $\varepsilon_{i}$.  A speckle distribution can be created experimentally by scattering laser light from a rough surface.  This results in a random potential that obeys the probability distribution, \mbox{$P(\varepsilon_{i})=(1/\langle \varepsilon_{i}\rangle)\exp (-\varepsilon_{i}/\langle \varepsilon_{i}\rangle)$}~\cite{Huntley,HorakCourtois} [see Figs.~\ref{figPotSamples}(c) and (f)] where $\langle.\rangle$ in this context denotes a mean.  We model the speckle numerically using the method outlined in Refs.~\cite{Huntley} and~\cite{HorakCourtois}.  Following recent experiments we define $\Delta=2\sigma_{SP}$, where $\sigma_{SP}$ is the standard deviation of $P(\varepsilon_{i})$~\cite{LyeFallani,SchulteDrenkelforth}.  For a speckle, $\langle\varepsilon_i\rangle=\sigma_{SP}$.  It should be noted that by using the BH Hamiltonian (\ref{eqBHH}), physically what is being modelled is not the speckle potential alone but the speckle plus an optical lattice.

Quasiperiodic disorder is different from the other types of disorder in that it is not random but rather, completely deterministic.  It can be created by superimposing two \emph{incommensurate} lattices (that is, the ratio of the periodicities is irrational).  Assuming that one lattice is much weaker than the other, the weaker lattice can be considered a disordered perturbation that shifts the well-depths of the first lattice and hence $\varepsilon_{i}$ by at most $\Delta$ [see Fig.~\ref{figPotSamples}(a)].  Under these circumstances the quasiperiodic potential can be represented in the BH Hamiltonian (\ref{eqBHH}) by $\varepsilon_{i}=\Delta\sin^{2}(kx_{i})$ where $k$ is irrational and $x_i=i\pi$.  $i$ here enumerates the lattice sites.  Comparing quasiperiodic disorder to random disorder is important as recent experiments have used quasiperiodic disorder to approximate uniform random disorder~\cite{FallaniLye}.

\begin{figure}
\includegraphics[width=8.7cm]{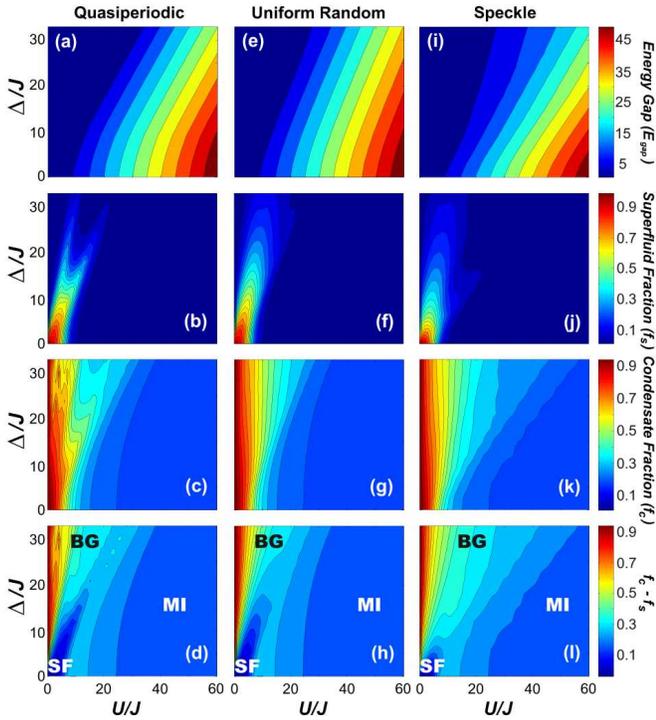}
\caption{\label{figContour}(Color online) The energy gap $E_{gap}$, the superfluid fraction $f_{s}$, the condensate fraction $f_{c}$, and $(f_{c}-f_{s})$, for the three different types of disordered lattices in Fig.~\ref{figPotSamples}.  $\Delta/J$ is the strength of the disorder and $U/J$ is the interaction strength, both scaled by the tunnelling strength $J$.  $N=7$ and $N_{s}=7$ for each sample.  For the uniform random disorder and the speckle-type disorder the results shown are the average of 24 samples.  BG=`Bose glass'; MI=`Mott insulator'; SF=`superfluid'.}
\end{figure}

Fig.~\ref{figContour} shows how $E_{gap}$, $f_{c}$, $f_{s}$ and $(f_{c}-f_{s})$ change with: (1) the type of disorder, (2) the strength of the disorder $\Delta$, and (3) the interatomic interactions $U$.  $\Delta$ and $U$ are both scaled by the tunnelling strength $J$.  $N=7$ and $N_s=7$ for each sample and for the two random types of disorder the average of 24 samples is taken.  For the quasiperiodic disorder we use $k=(1+\sqrt{5})/2$.

The zero disorder line ($\Delta/J=0$) represents the homogeneous periodic lattice.  As $U/J$ increases, the system undergoes a SF-MI (superfluid-Mott insulator) transition.  The MI is characterized here by: integer filling $\bar{n}_{i}$ (for our parameters $\bar{n}_{i}=1$), suppression of the number fluctuations $\sigma_{i}$, a finite energy gap, $f_{s}=0$ and $f_{c}$ going to $1/N_s$.

In the non-interacting case ($U/J=0$), disorder causes the atoms to become Anderson localized.  This is characterised by the exponential spatial localization of the atoms and $f_{s}=0$. $f_{c}$ is trivially equal to one here. 

\begin{figure}
\includegraphics[width=8.5cm]{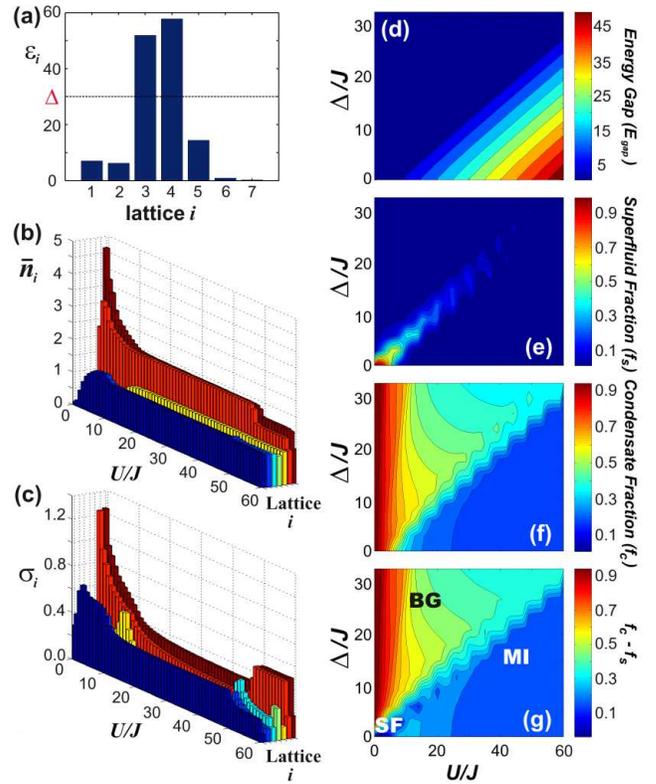}
\caption{\label{figSampleSpecklePot} (Color online) One of the speckle samples used in Fig.~\ref{figContour}.  (a)-(c) $\Delta/J=30$ only. (a) The topology of the disorder, $\varepsilon_i$.  The dotted line shows $\Delta/J$.  (b) number densities $\bar{n}_{i}$, (c) number fluctuations $\sigma_i$, (d) the energy gap $E_{gap}$, (e) the condensate fraction $f_c$, (f) the superfluid fraction $f_s$ and (g) the difference between the condensate and superfluid fractions.}
\end{figure}

For all three types of disorder and for any $\Delta/J$, as $U/J$ increases the onsite interactions eventually dominate and the system becomes a MI.  However, for strong enough disorder and moderate interactions $f_s$ is still zero or close to zero but the other properties of the state are different from that of a MI in a homogeneous lattice.   One difference is in the $\bar{n}_{i}$ [e.g. see Fig.~\ref{figSampleSpecklePot}(b)].  However, the most clear difference can be seen in the energy gap [Figs.~\ref{figContour}(a), (e) and (i)].  A Bose glass in a disordered interacting boson system is a \emph{gapless} compressible insulator~\cite{FisherWeichman}.  In comparison a MI is an incompressible insulator \emph{with a gap} which occurs for large enough $U/J$ and commensurate filling~\cite{FisherWeichman}.  A finite size system always has a discrete excitation spectrum and hence it is impossible to have $E_{gap}=0$.  However, in this region $E_{gap}$ is very small and many times weaker than the energy gap in the MI which reaches values of $E_{gap}\sim50$ in Fig.~\ref{figContour}.  In the MI, $E_{gap}$ increases linearly with $U/J$ so this is not likely to be its maximum value.  Thus in the transition from the Mott insulator to this region the energy gap can be said to effectively disappear.  Hence this region matches Fisher~\etal's description of the Bose glass~\cite{FisherWeichman}.

As Figs.~\ref{figContour}(d), (h) and (l) show, in both the superfluid and the MI regions there are no significant differences between $f_{c}$ and $f_{s}$.  However in the BG region it is clear that though $f_{c}$ is depleted it is significantly greater than $f_{s}$.  In the BG, \emph{there  exists condensate without superfluidity}.  Close to the MI or the superfluid regions, $[f_{c}-f_{s}]$ gradually decreases.  Therefore, condensate without superfluidity can be identified with the Bose glass.

\emph{Both} repulsive interactions and disorder are necessary for BG.  However, if $U/J\gg\Delta/J$ the system becomes a MI.  As can be seen in Fig.~\ref{figContour}, as $\Delta$ increases the critical $U$ required for the BG-MI quantum phase transition increases.  This means that in disordered lattices using very strong disorder will increase the size of the parameter regime where it is possible to obtain BG and hence condensate without superfluidity.  

The results above apply to all three types of disorder studied.  However significant differences exist between the different types of disorder.  First, the changes in $f_{s}$ and $f_{c}$ have a distinct complex lobe-like structure in the quasiperiodic case.  This has been observed previously for quasiperiodic disorder and attributed to the non-random nature of the quasiperiodic distribution and the sensitivity of rearrangements of the many-body wavefunction to the distribution of $\varepsilon_i$~\cite{RothBurnett,Bar-GillPugatch}. 

\begin{figure}
\includegraphics[width=8.6cm]{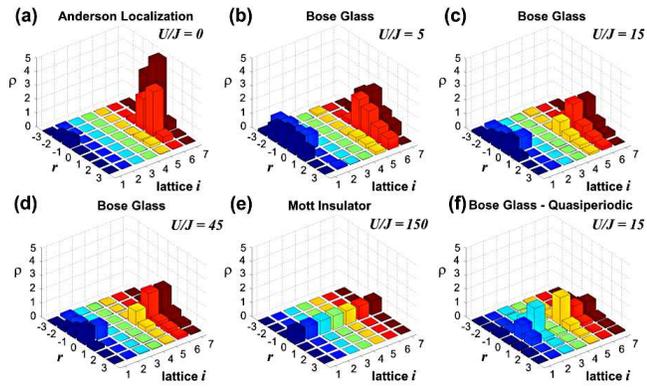}
\caption{\label{figDensityMatrix} (Color online) The spatial correlations $\rho(i,r)$ between sites $i$ and $i+r$ (with periodic boundary conditions).  $\rho(i,r=0)$ shows $\bar{n}_i$.  (a)-(e) is for the speckle sample shown in Fig.~\ref{figSampleSpecklePot}. (f) is for the quasiperiodic lattice in Fig.~\ref{figContour}.  All examples are for $\Delta/J=30$.}
\end{figure}

Another major difference exists between bounded (quasiperiodic and uniform random) and unbounded (speckle) disorder.  Fig.~\ref{figContour} shows that for the same $\Delta/J$ using unbounded rather than bounded disorder greatly increases the size of the BG region.  Fig.~\ref{figContour} shows the average over many samples. Examining individual speckle samples shows the largest BG regions occur when there are ``spikes" where $\varepsilon_{i} > \Delta$.  These inevitably occur due to the unbounded nature of the speckle distribution. An example of one such speckle sample with a ``spike" is given in Fig.~\ref{figSampleSpecklePot} which clearly shows the enlarged BG region.

Studying individual samples also allows us to examine the spatial correlations which are given by the off-diagonal terms of the one-body density matrix, \mbox{$\rho_{ij}=\langle\psi_{0}|\hat{a}_{j}^{\dagger}\hat{a}_{i}|\psi_{0}\rangle$}.  We let $j=i+r$ and use periodic boundary conditions;  $\rho(i,r=0)$ is thus simply $\bar{n}_{i}$.  Figs.~\ref{figDensityMatrix}(a)-(e) show $\rho(i,r)$ for the sample disorder in Fig.~\ref{figSampleSpecklePot}.  Fig.~\ref{figDensityMatrix}(f) is for the quasiperiodic disorder in Fig.~\ref{figContour}.  All examples are for $\Delta/J=30$.

In the non-interacting case [Fig.~\ref{figDensityMatrix}(a)] (Anderson localization), the wavefunction is strongly localized.  For $U/J\gg\Delta/J$ (Mott insulator), $\bar{n}_{i}=1$ for every lattice site regardless of $\varepsilon_{i}$ [Fig.~\ref{figDensityMatrix}(e)].  No correlations exist between lattice sites; each site is independent.

In the BG, the density distribution is much more homogeneous compared to the noninteracting case.  Unlike the MI, the BG retains significant correlations between lattice sites.  As $r$ increases $\rho(i,r)$ decreases but remains significant over a finite area.  This also occurs in other disorder samples though the exact details depend greatly on the topology of the sample.  Fig.~\ref{figDensityMatrix}(f) shows an example for quasiperiodic disorder.  Since $f_c$ is significantly greater than $f_s$ in the BG region, it is possible to consider this a localized homogeneous condensate pinned by the disorder.  The identification with the BG hints at the possible structure of the quantum phase though it remains to be seen if these results apply to larger systems.

As $U/J$ increases the spatial correlations between sites disappears.  However it is possible that in a true speckle potential (without a lattice) the increasing fragmentation of the condensate which can be observed here with increasing interaction strength may not occur due to the absence of a MI phase.  An alternative is using noninteger $N/N_s$ as previously studied in disordered lattices~\cite{RothBurnett,DamskiZakrzewski}.  

In conclusion we have shown how superfluid and condensate properties change in `dirty' Bose gases as interatomic interactions, disorder strength and the type of disorder are varied.  We showed that in a disordered lattice, the Bose glass as identified by the lack of superfluidity and energy gap, comprises of a BEC without superfluidity. In order to maximise the size of the parameter regime where it is possible to obtain BEC which is part of the normal fluid either the strength of the disorder can be increased or an unbounded disorder can be used.  By examining the spatial correlations between lattice sites in the different insulating states we found that the Bose glass may be considered a localized homogeneous condensate pinned by disorder. 

\section*{ACKNOWLEDGEMENTS}
We would like to acknowledge the support of the JSPS Postdoctoral Fellowship program and also the JSPS Grant-in-Aid for Scientific Research (Grant No. 18340109) and the MEXT Grant-in-Aid for Scientific Research on Priority Areas (Grant No. 17071008).


\begin{thebibliography}{18}
\expandafter\ifx\csname natexlab\endcsname\relax\def\natexlab#1{#1}\fi
\expandafter\ifx\csname bibnamefont\endcsname\relax
  \def\bibnamefont#1{#1}\fi
\expandafter\ifx\csname bibfnamefont\endcsname\relax
  \def\bibfnamefont#1{#1}\fi
\expandafter\ifx\csname citenamefont\endcsname\relax
  \def\citenamefont#1{#1}\fi
\expandafter\ifx\csname url\endcsname\relax
  \def\url#1{\texttt{#1}}\fi
\expandafter\ifx\csname urlprefix\endcsname\relax\def\urlprefix{URL }\fi
\providecommand{\bibinfo}[2]{#2}
\providecommand{\eprint}[2][]{\url{#2}}

\bibitem[{\citenamefont{Anderson}(1958)}]{Anderson}
\bibinfo{author}{\bibfnamefont{P.~W.} \bibnamefont{Anderson}},
  \bibinfo{journal}{Phys. Rev.} \textbf{\bibinfo{volume}{109}},
  \bibinfo{pages}{1492} (\bibinfo{year}{1958}).

\bibitem[{\citenamefont{Fisher et~al.}(1989)\citenamefont{Fisher, Weichman,
  Grinstein, and Fisher}}]{FisherWeichman}
\bibinfo{author}{\bibfnamefont{M.~P.~A.} \bibnamefont{Fisher}},
  \bibinfo{author}{\bibfnamefont{P.~B.} \bibnamefont{Weichman}},
  \bibinfo{author}{\bibfnamefont{G.}~\bibnamefont{Grinstein}},
  \bibnamefont{and} \bibinfo{author}{\bibfnamefont{D.~S.}
  \bibnamefont{Fisher}}, \bibinfo{journal}{Phys. Rev. B}
  \textbf{\bibinfo{volume}{40}}, \bibinfo{pages}{546} (\bibinfo{year}{1989}).

\bibitem[{\citenamefont{Huang and Meng}(1992)}]{HuangMeng}
\bibinfo{author}{\bibfnamefont{K.}~\bibnamefont{Huang}} \bibnamefont{and}
  \bibinfo{author}{\bibfnamefont{H.-F.} \bibnamefont{Meng}},
  \bibinfo{journal}{Phys. Rev. Lett.} \textbf{\bibinfo{volume}{69}},
  \bibinfo{pages}{644} (\bibinfo{year}{1992}).

\bibitem[{\citenamefont{Giorgini et~al.}(1994)\citenamefont{Giorgini,
  Pitaevskii, and Stringari}}]{GiorginiPitaevskii}
\bibinfo{author}{\bibfnamefont{S.}~\bibnamefont{Giorgini}},
  \bibinfo{author}{\bibfnamefont{L.}~\bibnamefont{Pitaevskii}},
  \bibnamefont{and}
  \bibinfo{author}{\bibfnamefont{S.}~\bibnamefont{Stringari}},
  \bibinfo{journal}{Phys. Rev. B} \textbf{\bibinfo{volume}{49}},
  \bibinfo{pages}{12938} (\bibinfo{year}{1994}).

\bibitem[{\citenamefont{Damski et~al.}(2003)\citenamefont{Damski, Zakrzewski,
  Santos, Zoller, and Lewenstein}}]{DamskiZakrzewski}
\bibinfo{author}{\bibfnamefont{B.}~\bibnamefont{Damski}},
  \bibinfo{author}{\bibfnamefont{J.}~\bibnamefont{Zakrzewski}},
  \bibinfo{author}{\bibfnamefont{L.}~\bibnamefont{Santos}},
  \bibinfo{author}{\bibfnamefont{P.}~\bibnamefont{Zoller}}, \bibnamefont{and}
  \bibinfo{author}{\bibfnamefont{M.}~\bibnamefont{Lewenstein}},
  \bibinfo{journal}{Phys. Rev. Lett.} \textbf{\bibinfo{volume}{91}},
  \bibinfo{pages}{080403} (\bibinfo{year}{2003}).

\bibitem[{\citenamefont{Roth and Burnett}(2003)}]{RothBurnett}
\bibinfo{author}{\bibfnamefont{R.}~\bibnamefont{Roth}} \bibnamefont{and}
  \bibinfo{author}{\bibfnamefont{K.}~\bibnamefont{Burnett}},
  \bibinfo{journal}{Phys. Rev. A} \textbf{\bibinfo{volume}{68}},
  \bibinfo{pages}{023604} (\bibinfo{year}{2003}).

\bibitem[{\citenamefont{Reppy}(1992)}]{Reppy}
\bibinfo{author}{\bibfnamefont{J.~D.} \bibnamefont{Reppy}},
  \bibinfo{journal}{J. Low Temp. Phys.} \textbf{\bibinfo{volume}{87}},
  \bibinfo{pages}{205} (\bibinfo{year}{1992}).

\bibitem[{\citenamefont{Lye et~al.}(2005)\citenamefont{Lye, Fallani
  et~al.}}]{LyeFallani}
\bibinfo{author}{\bibfnamefont{J.~E.} \bibnamefont{Lye}},
  \bibinfo{author}{\bibfnamefont{L.}~\bibnamefont{Fallani}},
  \bibnamefont{et~al.}, \bibinfo{journal}{Phys. Rev. Lett.}
  \textbf{\bibinfo{volume}{95}}, \bibinfo{pages}{070401}
  (\bibinfo{year}{2005}).

\bibitem[{\citenamefont{Cl{\'e}ment et~al.}(2005)\citenamefont{Cl{\'e}ment,
  V{\'a}ron et~al.}}]{ClementVaron}
\bibinfo{author}{\bibfnamefont{D.}~\bibnamefont{Cl{\'e}ment}},
  \bibinfo{author}{\bibfnamefont{A.~F.} \bibnamefont{V{\'a}ron}},
  \bibnamefont{et~al.}, \bibinfo{journal}{Phys. Rev. Lett.}
  \textbf{\bibinfo{volume}{95}}, \bibinfo{pages}{170409}
  (\bibinfo{year}{2005}).

\bibitem[{\citenamefont{Schulte et~al.}(2005)\citenamefont{Schulte,
  Drenkelforth et~al.}}]{SchulteDrenkelforth}
\bibinfo{author}{\bibfnamefont{T.}~\bibnamefont{Schulte}},
  \bibinfo{author}{\bibfnamefont{S.}~\bibnamefont{Drenkelforth}},
  \bibnamefont{et~al.}, \bibinfo{journal}{Phys. Rev. Lett.}
  \textbf{\bibinfo{volume}{95}}, \bibinfo{pages}{170411}
  (\bibinfo{year}{2005}).

\bibitem[{\citenamefont{Fallani et~al.}()\citenamefont{Fallani, Lye, Guarrera,
  Fort, and Inguscio}}]{FallaniLye}
\bibinfo{author}{\bibfnamefont{L.}~\bibnamefont{Fallani}},
  \bibinfo{author}{\bibfnamefont{J.~E.} \bibnamefont{Lye}},
  \bibinfo{author}{\bibfnamefont{V.}~\bibnamefont{Guarrera}},
  \bibinfo{author}{\bibfnamefont{C.}~\bibnamefont{Fort}}, \bibnamefont{and}
  \bibinfo{author}{\bibfnamefont{M.}~\bibnamefont{Inguscio}},
  \bibinfo{note}{cond-mat/0603655 (2006)}.

\bibitem[{\citenamefont{Astrakharchik et~al.}(2002)\citenamefont{Astrakharchik,
  Boronat, Casulleras, and Giorgini}}]{AstrakharchikBoronat}
\bibinfo{author}{\bibfnamefont{G.~E.} \bibnamefont{Astrakharchik}},
  \bibinfo{author}{\bibfnamefont{J.}~\bibnamefont{Boronat}},
  \bibinfo{author}{\bibfnamefont{J.}~\bibnamefont{Casulleras}},
  \bibnamefont{and} \bibinfo{author}{\bibfnamefont{S.}~\bibnamefont{Giorgini}},
  \bibinfo{journal}{Phys. Rev. A} \textbf{\bibinfo{volume}{66}},
  \bibinfo{pages}{023603} (\bibinfo{year}{2002}).

\bibitem[{\citenamefont{Jaksch et~al.}(1998)\citenamefont{Jaksch, Bruder,
  Cirac, Gardiner, and Zoller}}]{JakschBruder}
\bibinfo{author}{\bibfnamefont{D.}~\bibnamefont{Jaksch}},
  \bibinfo{author}{\bibfnamefont{C.}~\bibnamefont{Bruder}},
  \bibinfo{author}{\bibfnamefont{J.~I.} \bibnamefont{Cirac}},
  \bibinfo{author}{\bibfnamefont{C.~W.} \bibnamefont{Gardiner}},
  \bibnamefont{and} \bibinfo{author}{\bibfnamefont{P.}~\bibnamefont{Zoller}},
  \bibinfo{journal}{Phys. Rev. Lett.} \textbf{\bibinfo{volume}{81}},
  \bibinfo{pages}{3108} (\bibinfo{year}{1998}).

\bibitem[{\citenamefont{Penrose and Onsager}(1956)}]{PenroseOnsager}
\bibinfo{author}{\bibfnamefont{O.}~\bibnamefont{Penrose}} \bibnamefont{and}
  \bibinfo{author}{\bibfnamefont{L.}~\bibnamefont{Onsager}},
  \bibinfo{journal}{Phys. Rev.} \textbf{\bibinfo{volume}{104}},
  \bibinfo{pages}{576} (\bibinfo{year}{1956}).

\bibitem[{\citenamefont{Huntley}(1989)}]{Huntley}
\bibinfo{author}{\bibfnamefont{J.~M.} \bibnamefont{Huntley}},
  \bibinfo{journal}{Appl. Opt.} \textbf{\bibinfo{volume}{28}},
  \bibinfo{pages}{4316} (\bibinfo{year}{1989}).

\bibitem[{\citenamefont{Horak et~al.}(1998)\citenamefont{Horak, Courtois, and
  Grynberg}}]{HorakCourtois}
\bibinfo{author}{\bibfnamefont{P.}~\bibnamefont{Horak}},
  \bibinfo{author}{\bibfnamefont{J.-Y.} \bibnamefont{Courtois}},
  \bibnamefont{and} \bibinfo{author}{\bibfnamefont{G.}~\bibnamefont{Grynberg}},
  \bibinfo{journal}{Phys. Rev. A} \textbf{\bibinfo{volume}{58}},
  \bibinfo{pages}{3953} (\bibinfo{year}{1998}).

\bibitem[{\citenamefont{Bar-Gill et~al.}()\citenamefont{Bar-Gill, Pugatch,
  Rowen, Katz, and Davidson}}]{Bar-GillPugatch}
\bibinfo{author}{\bibfnamefont{N.}~\bibnamefont{Bar-Gill}},
  \bibinfo{author}{\bibfnamefont{R.}~\bibnamefont{Pugatch}},
  \bibinfo{author}{\bibfnamefont{E.}~\bibnamefont{Rowen}},
  \bibinfo{author}{\bibfnamefont{N.}~\bibnamefont{Katz}}, \bibnamefont{and}
  \bibinfo{author}{\bibfnamefont{N.}~\bibnamefont{Davidson}},
  \bibinfo{note}{cond-mat/0603513 (2006)}.

\end{thebibliography}
\end{document}